\pdfoutput=1
\documentclass[12pt,a4paper]{article}

\usepackage{ifthen} 
\newboolean{pdflatex}
\setboolean{pdflatex}{true} 

\newboolean{articletitles}
\setboolean{articletitles}{true} 

\newboolean{uprightparticles}
\setboolean{uprightparticles}{false} 

\newcommand{\X}{\chi_{c1}(3872)}

\newcommand{\N}{\mathrm{N_{tracks}^{VELO}}}

\def\paperauthors{LHCb collaboration} 
\def\paperasciititle{Observation of multiplicity-dependent prompt chi{c1}(3872) and psi(2S) production in pp collisions} 
\def\papertitle{Observation of multiplicity-dependent prompt $\chi_{c1}(3872)$ and $\psitwos$ production\\ in $pp$ collisions} 

\def\paperkeywords{{LHCb}} 
\def\papercopyright{\the\year\ CERN for the benefit of the LHCb collaboration} 
\def\paperlicence{CC BY 4.0 licence}
\def\paperlicenceurl{https://creativecommons.org/licenses/by/4.0/}

\usepackage[top=1in, bottom=1.25in, left=1in, right=1in]{geometry}

\usepackage{microtype}
\usepackage{lineno}  
\usepackage{xspace} 
\usepackage{caption} 

\usepackage{graphicx}  
\usepackage{color}
\usepackage{colortbl}
\graphicspath{{./figs/}} 

\usepackage{amsmath} 
\usepackage{amssymb}
\usepackage{amsfonts}
\usepackage{upgreek} 

\newcommand*\patchAmsMathEnvironmentForLineno[1]{%
\expandafter\let\csname old#1\expandafter\endcsname\csname #1\endcsname
\expandafter\let\csname oldend#1\expandafter\endcsname\csname
end#1\endcsname
 \renewenvironment{#1}%
   {\linenomath\csname old#1\endcsname}%
   {\csname oldend#1\endcsname\endlinenomath}%
}
\newcommand*\patchBothAmsMathEnvironmentsForLineno[1]{%
  \patchAmsMathEnvironmentForLineno{#1}%
  \patchAmsMathEnvironmentForLineno{#1*}%
}
\AtBeginDocument{%
\patchBothAmsMathEnvironmentsForLineno{equation}%
\patchBothAmsMathEnvironmentsForLineno{align}%
\patchBothAmsMathEnvironmentsForLineno{flalign}%
\patchBothAmsMathEnvironmentsForLineno{alignat}%
\patchBothAmsMathEnvironmentsForLineno{gather}%
\patchBothAmsMathEnvironmentsForLineno{multline}%
\patchBothAmsMathEnvironmentsForLineno{eqnarray}%
}

\usepackage{hyperxmp}

\usepackage[pdftex,
            pdfauthor={\paperauthors},
            pdftitle={\paperasciititle},
            pdfkeywords={\paperkeywords},
            pdfcopyright={Copyright (C) \papercopyright},
            pdflicenseurl={\paperlicenceurl}]{hyperref}

\usepackage[colorinlistoftodos,textsize=scriptsize]{todonotes}

\usepackage[bottom,flushmargin,hang,multiple]{footmisc}

\usepackage[all]{hypcap} 

\usepackage{xspace} 
\usepackage{upgreek}

\def\lhcb   {\mbox{LHCb}\xspace}

\def\belle  {\mbox{Belle}\xspace}

\def\MagUp {\mbox{\em Mag\kern -0.05em Up}\xspace}

\ifthenelse{\boolean{uprightparticles}}%
{

 \def\Pmu         {\ensuremath{\upmu}\xspace}

 \def\Ppi         {\ensuremath{\uppi}\xspace}

 \def\Pchi        {\ensuremath{\upchi}\xspace}                 
 \def\Ppsi        {\ensuremath{\uppsi}\xspace}

 \def\PDelta      {\ensuremath{\Delta}\xspace}                 
 \def\PXi         {\ensuremath{\Xi}\xspace}                 
 \def\PLambda     {\ensuremath{\Lambda}\xspace}                 
 \def\PSigma      {\ensuremath{\Sigma}\xspace}                 
 \def\POmega      {\ensuremath{\Omega}\xspace}                 
 \def\PUpsilon    {\ensuremath{\Upsilon}\xspace}

 \def\PB      {\ensuremath{\mathrm{B}}\xspace}                 
                  
 \def\PD      {\ensuremath{\mathrm{D}}\xspace}

 \def\PJ      {\ensuremath{\mathrm{J}}\xspace}                 
 \def\PK      {\ensuremath{\mathrm{K}}\xspace}

 \def\Pb      {\ensuremath{\mathrm{b}}\xspace}                 
 \def\Pc      {\ensuremath{\mathrm{c}}\xspace}

 \def\Pi      {\ensuremath{\mathrm{i}}\xspace}

 \def\Ps      {\ensuremath{\mathrm{s}}\xspace}

 \def\thebaroffset{0.0em}
}
{

 \def\Pmu         {\ensuremath{\mu}\xspace}

 \def\Ppi         {\ensuremath{\pi}\xspace}

 \def\Pchi        {\ensuremath{\chi}\xspace}                 
 \def\Ppsi        {\ensuremath{\psi}\xspace}                 
                  
 \mathchardef\PDelta="7101
 \mathchardef\PXi="7104
 \mathchardef\PLambda="7103
 \mathchardef\PSigma="7106
 \mathchardef\POmega="710A
 \mathchardef\PUpsilon="7107
                  
 \def\PB      {\ensuremath{B}\xspace}                 
                  
 \def\PD      {\ensuremath{D}\xspace}

 \def\PJ      {\ensuremath{J}\xspace}                 
 \def\PK      {\ensuremath{K}\xspace}

 \def\Pb      {\ensuremath{b}\xspace}                 
 \def\Pc      {\ensuremath{c}\xspace}

 \def\Pi      {\ensuremath{i}\xspace}

 \def\Ps      {\ensuremath{s}\xspace}

 \def\thebaroffset{0.18em}
}
\newcommand{\offsetoverline}[2][\thebaroffset]{\kern #1\overline{\kern -#1 #2}}%

\makeatletter
\ifcase \@ptsize \relax
  \newcommand{\miniscule}{\@setfontsize\miniscule{4}{5}}
\or
  \newcommand{\miniscule}{\@setfontsize\miniscule{5}{6}}
\or
  \newcommand{\miniscule}{\@setfontsize\miniscule{5}{6}}
\fi
\makeatother

\DeclareRobustCommand{\optbar}[1]{\shortstack{{\miniscule (\rule[.5ex]{1.25em}{.18mm})}
  \\ [-.7ex] $#1$}}

\def\mup        {{\ensuremath{\Pmu^+}}\xspace}

\def\squark    {{\ensuremath{\Ps}}\xspace}

\def\cquark    {{\ensuremath{\Pc}}\xspace}

\def\bquark    {{\ensuremath{\Pb}}\xspace}

\def\pion   {{\ensuremath{\Ppi}}\xspace}

\def\pip    {{\ensuremath{\pion^+}}\xspace}
\def\pim    {{\ensuremath{\pion^-}}\xspace}

\def\KorKbar {\kern \thebaroffset\optbar{\kern -\thebaroffset \PK}{}\xspace}

\def\Dbar    {{\ensuremath{\offsetoverline{\PD}}}\xspace}
\def\D       {{\ensuremath{\PD}}\xspace}

\def\DorDbar {\kern \thebaroffset\optbar{\kern -\thebaroffset \PD}\xspace}
\def\Dz      {{\ensuremath{\D^0}}\xspace}

\def\Dp      {{\ensuremath{\D^+}}\xspace}
\def\Dm      {{\ensuremath{\D^-}}\xspace}

\def\DpDm    {\ensuremath{\Dp {\kern -0.16em \Dm}}\xspace}

\def\Dstarzb {{\ensuremath{\Dbar{}^{*0}}}\xspace}

\def\B       {{\ensuremath{\PB}}\xspace}

\def\BorBbar {\kern \thebaroffset\optbar{\kern -\thebaroffset \PB}\xspace}

\def\Bd      {{\ensuremath{\B^0}}\xspace}

\def\BdorBdbar {\kern \thebaroffset\optbar{\kern -\thebaroffset \Bd}\xspace}

\def\Bs      {{\ensuremath{\B^0_\squark}}\xspace}

\def\BsorBsbar {\kern \thebaroffset\optbar{\kern -\thebaroffset \Bs}\xspace}

\def\jpsi     {{\ensuremath{{\PJ\mskip -3mu/\mskip -2mu\Ppsi}}}\xspace}
\def\psitwos  {{\ensuremath{\Ppsi{(2S)}}}\xspace}

\def\Upsilonres  {{\ensuremath{\PUpsilon}}\xspace}
\def\Y#1S{\ensuremath{\PUpsilon{(#1S)}}\xspace}
\def\OneS  {{\Y1S}}
\def\TwoS  {{\Y2S}}
\def\ThreeS{{\Y3S}}

\def\theX     {{\ensuremath{\Pchi_{c1}(3872)}}\xspace}

\def\LorLbar     {\kern \thebaroffset\optbar{\kern -\thebaroffset \PLambda}\xspace}

\def\BF         {{\ensuremath{\mathcal{B}}}\xspace}
\def\BR         {\BF}

\def\AT#1     {\ensuremath{A_{\mathrm{T}}^{#1}}\xspace}           

\def\C#1      {\ensuremath{\mathcal{C}_{#1}}\xspace}                       
\def\Cp#1     {\ensuremath{\mathcal{C}_{#1}^{'}}\xspace}                    
\def\Ceff#1   {\ensuremath{\mathcal{C}_{#1}^{\mathrm{(eff)}}}\xspace}        
\def\Cpeff#1  {\ensuremath{\mathcal{C}_{#1}^{'\mathrm{(eff)}}}\xspace}       
\def\Ope#1    {\ensuremath{\mathcal{O}_{#1}}\xspace}                       
\def\Opep#1   {\ensuremath{\mathcal{O}_{#1}^{'}}\xspace}                    

\newcommand{\aunit}[1]{\ensuremath{\text{\,#1}}}       

\newcommand{\tev}{\aunit{Te\kern -0.1em V}\xspace}
\newcommand{\gev}{\aunit{Ge\kern -0.1em V}\xspace}
\newcommand{\mev}{\aunit{Me\kern -0.1em V}\xspace}
\newcommand{\kev}{\aunit{ke\kern -0.1em V}\xspace}
\newcommand{\ev}{\aunit{e\kern -0.1em V}\xspace}
 
\newcommand{\mevc}{\ensuremath{\aunit{Me\kern -0.1em V\!/}c}\xspace}
\newcommand{\gevc}{\ensuremath{\aunit{Ge\kern -0.1em V\!/}c}\xspace}
\newcommand{\mevcc}{\ensuremath{\aunit{Me\kern -0.1em V\!/}c^2}\xspace}
\newcommand{\gevcc}{\ensuremath{\aunit{Ge\kern -0.1em V\!/}c^2}\xspace}

\def\mm   {\aunit{mm}\xspace}

\def\fm   {\aunit{fm}\xspace}

\def\fb   {\ensuremath{\aunit{fb}}\xspace}
\def\invfb   {\ensuremath{\fb^{-1}}\xspace}

\def\gsim{{~\raise.15em\hbox{$>$}\kern-.85em
          \lower.35em\hbox{$\sim$}~}\xspace}
\def\lsim{{~\raise.15em\hbox{$<$}\kern-.85em
          \lower.35em\hbox{$\sim$}~}\xspace}

\def\pt         {\ensuremath{p_{\mathrm{T}}}\xspace}

\def\ptot       {\ensuremath{p}\xspace}

\def\evtgen     {\mbox{\textsc{EvtGen}}\xspace}

\def\geant      {\mbox{\textsc{Geant4}}\xspace}

\def\pythia     {\mbox{\textsc{Pythia}}\xspace}

\def\tell1  {TELL1\xspace}
\def\ukl1   {UKL1\xspace}

\usepackage{cite} 
\usepackage{mciteplus}

\def\fprompt {\ensuremath{f_{\mathrm{prompt}}}\xspace}

\def\psitwos  {{\ensuremath{\Ppsi{(2S)}}}\xspace}

\begin{document}

\renewcommand{\thefootnote}{\fnsymbol{footnote}}
\setcounter{footnote}{1}

\onecolumn

\begin{titlepage}
\pagenumbering{roman}

\vspace*{-1.5cm}
\centerline{\large EUROPEAN ORGANIZATION FOR NUCLEAR RESEARCH (CERN)}
\vspace*{1.5cm}
\noindent
\begin{tabular*}{\linewidth}{lc@{\extracolsep{\fill}}r@{\extracolsep{0pt}}}
\ifthenelse{\boolean{pdflatex}}
{\vspace*{-1.5cm}\mbox{\!\!\!\includegraphics[width=.14\textwidth]{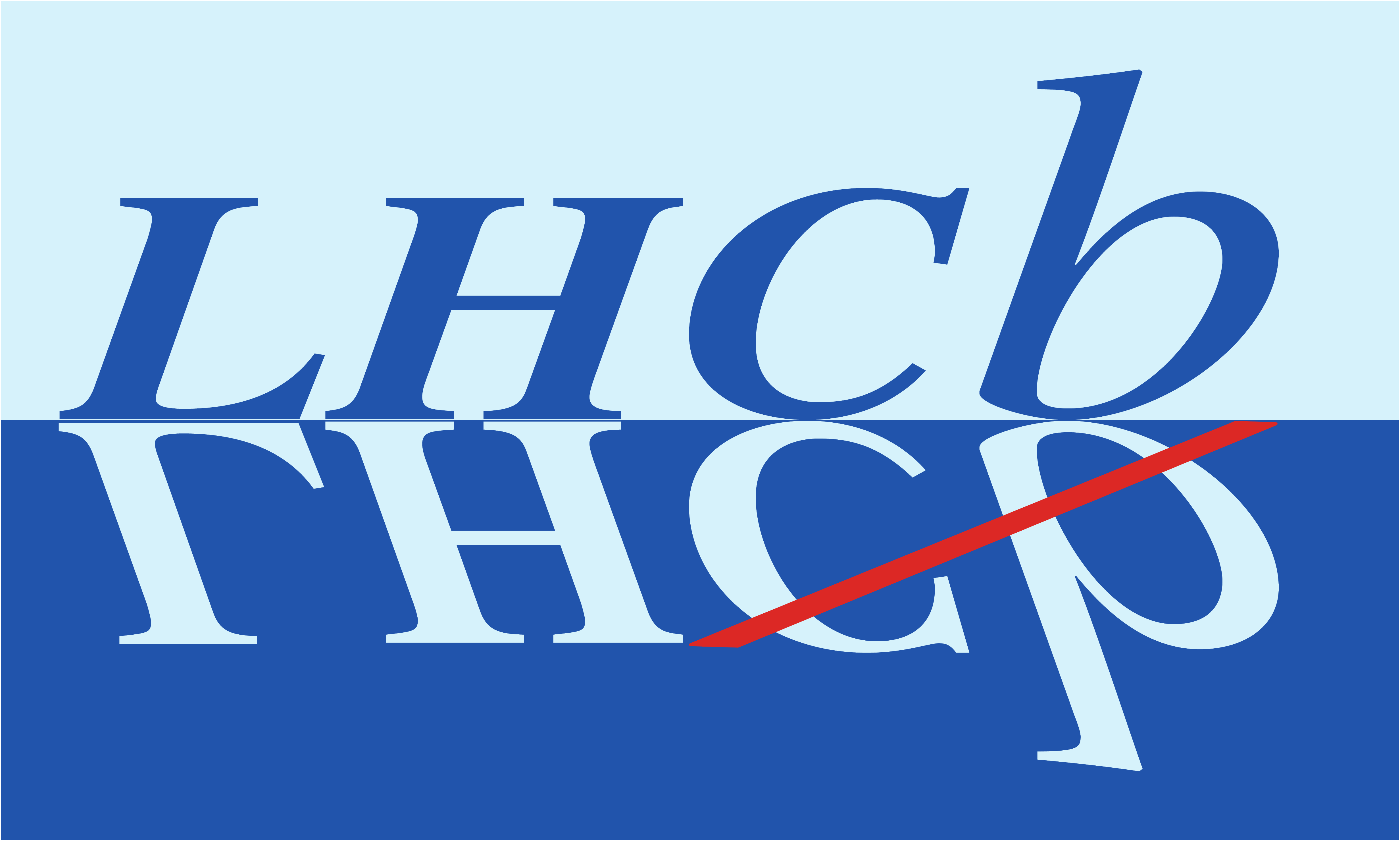}} & &}%
{\vspace*{-1.2cm}\mbox{\!\!\!\includegraphics[width=.12\textwidth]{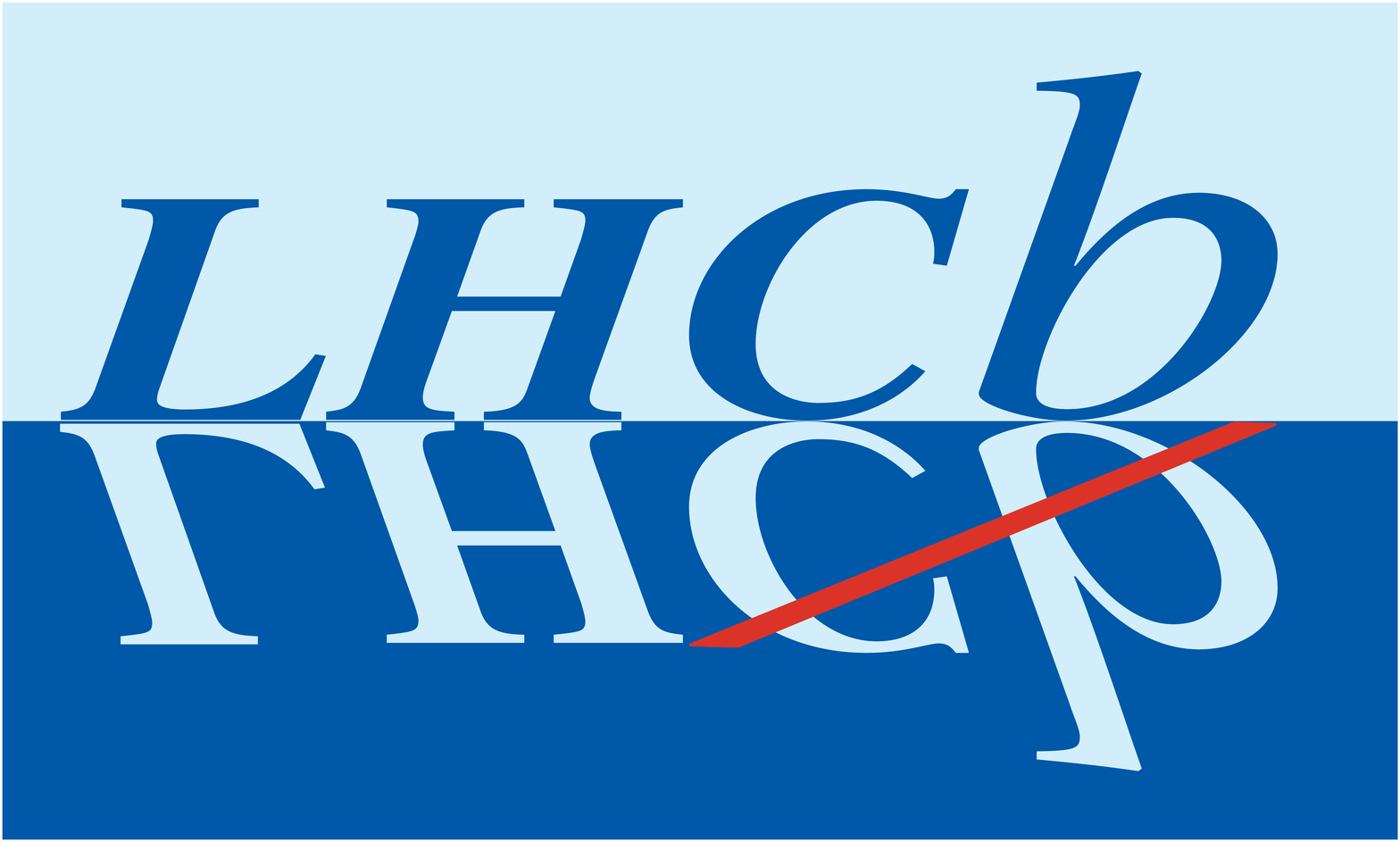}} & &}%
\\
 & & CERN-EP-2020-161 \\  
 & & LHCb-PAPER-2020-023 \\  
 & & 14 September 2020 \\ 
 & & \\
\end{tabular*}

\vspace*{4.0cm}

{\normalfont\bfseries\boldmath\huge
\begin{center}
  \papertitle 
\end{center}
}

\vspace*{2.0cm}

\begin{center}
\paperauthors\footnote{Authors are listed at the end of this Letter.}
\end{center}

\vspace{\fill}

\begin{abstract}
  \noindent
  The production of $\X$ and $\psitwos$ hadrons is studied as a function of charged particle multiplicity in $pp$ collisions at a center-of-mass energy of $8\tev$, corresponding to an integrated luminosity of 2\invfb.  For both states, the fraction that is produced promptly at the collision vertex is found to decrease as charged particle multiplicity increases. The ratio of $\X$ to $\psitwos$ cross-sections for promptly produced particles is also found to decrease with multiplicity, while no significant dependence on multiplicity is observed for the equivalent ratio of particles produced away from the collision vertex in $b$-hadron decays.  This behavior is consistent with a calculation that models the $\X$ structure as a compact tetraquark. Implications for the binding energy of the $\X$ state are discussed.  
  
\end{abstract}

\vspace*{1.8cm}

\begin{center}
  Published in
  Phys.~Rev.~Lett. 126 (2021) 092001
\end{center}

\vspace{\fill}

{\footnotesize 
\centerline{\copyright~\papercopyright. \href{\paperlicenceurl}{\paperlicence}.}}
\vspace*{2mm}

\end{titlepage}


\newpage
\setcounter{page}{2}
\mbox{~}
%
%
%
%


\renewcommand{\thefootnote}{\arabic{footnote}}
\setcounter{footnote}{0}

\cleardoublepage


\pagestyle{plain} 
\setcounter{page}{1}
\pagenumbering{arabic}


In recent years, multiple new resonances containing heavy quarks have been observed that do not fit into the framework of conventional hadrons, see Ref. \cite{ExoticReview} for a recent review.  The most studied of these exotic hadrons is the $\theX$ state, also known as $X(3872)$.  It was first discovered in the mass spectrum of $\jpsi \pip \pim$ in $B$-meson decays by the  \belle collaboration \cite{Belle_X3872obs}, and has since been confirmed by multiple other experiments \cite{CDF_X3872obs, D0_X3872obs, BABAR_X3872obs, LHCb-PAPER-2011-034}.  Despite intense scrutiny, the exact nature of the $\theX$ state is still unclear.

Multiple explanations of the $\X$ structure have been proposed. Shortly after its discovery, it was considered as one of several possible charmonium states \cite{XasCharmonium}. However, LHCb has since measured the quantum numbers to be $J^{PC}=1^{++}$ \cite{LHCb-PAPER-2013-001}, which disfavors its assignment as conventional charmonium because no compatible charmonium states with these quantum numbers are expected near the measured mass~\cite{CharmoniumPotentialModel}.  Other models consider the $\X$ state to be a tetraquark, which may have further substructure, composed of a diquark-antidiquark bound state~\cite{X3872asDiquark, X3872asDiquark2, X3872asTetraquark} or a hadrocharmonium state where two light quarks orbit a charmonium core~\cite{X3872asHadrocharmonium}.  Mixtures of various exotic and conventional states have also been studied~\cite{Colangelo:2007ph,X3872asMix, X3872asMix2, X3872Radius}.  The remarkable proximity of the $\X$ mass to the sum of the \Dz and \Dstarzb meson masses have led to the consideration of its structure as a hadronic molecule, a state comprising these two mesons bound via pion exchange~\cite{X3872asDD,X3872asDD2}.  In this case the binding energy of the $\X$ hadron would be small, as the mass difference $(M_{\Dz} + M_{\Dstarzb}) - M_{\X} = 0.07 \pm 0.12$\mevcc is consistent with zero \cite{LHCb-PAPER-2020-009}.  Consequently, these models assign the $\X$ state a large radius of $O$(10$\fm$)~\cite{X3872Radius, X3872Radius2}.  Results from recent LHCb studies of the $\X$ line shape are compatible with the molecular interpretation but do not exclude other possibilities \cite{LHCb-PAPER-2020-009,LHCb-PAPER-2020-008}.  

Techniques developed to study quarkonium production in proton-nucleus ($p$A) collisions can be used to probe the binding energy of hadrons.  Measurements of charmonium production in $p$A collisions at fixed target experiments~\cite{E866,NA50} and colliders~\cite{PPG151, ALICEpPb, LHCb-PAPER-2015-058, PPG188,CMSpPbpsi2s,ALICEpPb2} showed that $\psitwos$ production is suppressed more than $\jpsi$ production in rapidity regions where a relatively large number of charged particles are produced.  Similarly, measurements of $\Upsilonres$ production at the Large Hadron Collider (LHC) revealed that the $\TwoS$ and $\ThreeS$ states are suppressed more than the $\OneS$~\cite{CMSupsilon, LHCb-PAPER-2018-035}.  As the effects governing heavy quark production and transport through the nucleus are assumed to be similar for states with the same quark content, the mechanism for the suppression of excited states is expected to occur in the late stages of the collision, after the heavy quark pair has hadronized into a final state.  Models incorporating final-state effects, such as heavy quark pair breakup via interactions with comoving hadrons, describe the relative suppression of excited quarkonium states in $p$A collisions\cite{CapellaComovers, Capella:2000zp, Ferreiro_comovers, CGCcomover, Ferreiro2018}.  Similar final-state effects can also disrupt formation of the $\X$ state via interactions with pions produced in the underlying event \cite{BraatenBreakup}, and would be especially significant if the $\X$ structure is a large, weakly bound hadronic molecule.

High-multiplicity $pp$ collisions provide a hadronic environment that approaches heavy ion collisions in many respects.  Recently, phenomena typically thought only to occur in collisions of large nuclei have been observed in high-multiplicity $pp$ collisions, including a near-side ridge in two-particle angular correlations \cite{CMSppridge}, strangeness enhancement \cite{ALICEstrangeness}, and collective flow \cite{CMSppcollectivity}.  Multiplicity-dependent modification of $\Upsilonres$ production has also been observed \cite{Sirunyan:2020zzb}.
Therefore, high-multiplicity $pp$ collisions provide a testing ground for examining final-state effects observed on quarkonium in $p$A and AA collisions.  Measurements of such effects can provide new constraints on the structure of \mbox{the $\X$ \cite{Esposito:2020ywk}}.

In this Letter, measurements of the fractions of $\X$ and $\psitwos$ states, \fprompt, that are produced promptly at the $pp$ collision vertex as a function of charged particle multiplicity are presented.  The $\X$ and $\psitwos$ states are compared by measuring the ratio of the $\X$ to $\psitwos$ cross-sections, as a function of multiplicity.  The $\X$ and $\psitwos$ candidates are reconstructed through their decays to the $\jpsi\pip \pim$ final state, where the $\jpsi$ meson subsequently decays to $\mup \mu^-$ pairs.  This study uses data collected with the LHCb detector at a center-of-mass energy $\sqrt{s} = 8$\tev, corresponding to an integrated luminosity of 2\invfb.

The LHCb detector~\cite{LHCb-DP-2008-001,LHCb-DP-2014-002} is a
single-arm forward spectrometer covering the pseudorapidity range $2 < \eta < 5$, designed for the study of particles containing \bquark\ or \cquark\ quarks. The detector elements comprise a silicon-strip vertex detector (VELO) surrounding the $pp$ interaction region that allows \bquark\ hadrons to be identified from their characteristically long flight distance; a tracking system that provides a measurement of the momentum, $p$, of charged particles; two ring-imaging Cherenkov (RICH) detectors that discriminate between different species of charged hadrons, and a series of tracking detectors interleaved with hadron absorbers for identifying muons.  In this analysis, multiplicity is represented by the number of charged particle tracks reconstructed in the VELO, $\N$.  The VELO track-reconstruction efficiency has been measured to be about $99\%$~\cite{LHCB-DP-2014-001}. 

Simulation is required to model the effects of the detector acceptance and the imposed selection requirements. In the simulation, $pp$ collisions are generated using \pythia~\cite{Sjostrand:2007gs,Sjostrand:2006za} with a specific \lhcb configuration~\cite{LHCb-PROC-2010-056}.  Decays of unstable particles are described by \evtgen~\cite{Lange:2001uf}. The interaction of the generated particles with the detector, and its response, are implemented using the \geant toolkit~\cite{Allison:2006ve, *Agostinelli:2002hh} as described in Ref.~\cite{LHCb-PROC-2011-006}.  

Events considered in this analysis are selected by a set of triggers designed to record events containing the decay $\jpsi\rightarrow\mup\mu^-$. Tracks from triggered events that are identified as good muon candidates are retained. The muons are required to have momentum $\ptot> 10 \gevc$ and a momentum component transverse to the beam direction $\pt> 650 \mevc$.  Candidate $\jpsi$ mesons are formed from a pair of oppositely charged muons with an invariant mass within $\pm39 \mevcc$ (corresponding to 3 times the resolution on the mass) of the known $\jpsi$ mass and combined \mbox{$\pt> 3 \gevc$}. Charged pion candidates are selected using particle identification information from the RICH detectors. They are required to have $\ptot> 3 \gevc$ to ensure that the pions are above threshold in one of the RICH detectors, and have $\pt> 500 \mevc$ to reduce combinatorial background. 

\begin{figure}[b]
  \centering
 \includegraphics[width=1.0\linewidth]{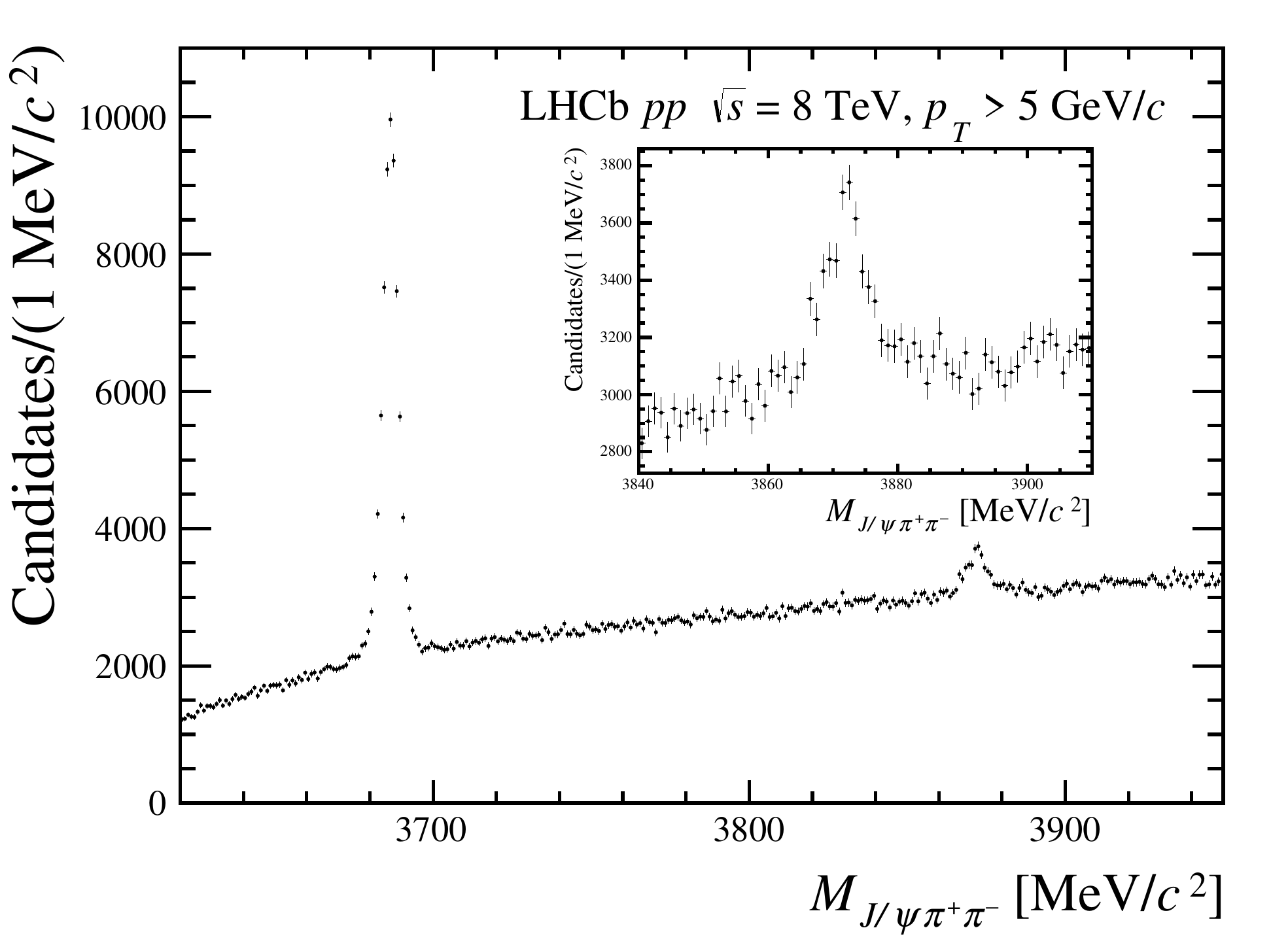}
  \caption{The $\jpsi \pi^+ \pi^-$ invariant-mass spectrum.  The inset shows the region of the $\X$ resonance.}
  \label{fig:mass}
\end{figure}

Selected $\mup\mu^-\pip\pim$ combinations that form a good-quality common vertex are fitted  with kinematic constraints that require all tracks to originate from a common vertex and constrain the dimuon mass to the known $\jpsi$ mass~\cite{PDG2018}.  The decay kinematics are required to satisfy $M_{\jpsi\pip\pim} - M_{\jpsi} - M_{\pip\pim} < 300 \mevcc$ and the candidates must have $\pt > 5 \gevc$ and be within the \mbox{pseudorapidity} range $2<\eta <4.5$.  The resulting $\jpsi\pip\pim$ invariant-mass spectrum is shown in Fig.~\ref{fig:mass}.

To avoid multiplicity biases arising from tracks produced in multiple collisions that occur in the same beam crossing, only events with a single reconstructed collision vertex are considered.  The position of collision vertices is restricted to a range along the beam direction  $-60 < z < 120 \mm$, to avoid biases from missing tracks that fall outside the VELO acceptance.

Both $\X$ and $\psitwos$ hadrons can be produced promptly at the $pp$ collision vertex, either directly or in strong decays of higher charmonia states, or in the decays of $b$ hadrons, which travel several millimeters before decaying.  The prompt component of the signal is separated from the component originating from $b$ decays by performing a simultaneous fit to the $\jpsi\pip\pim$ invariant-mass spectrum and the pseudo decay-time spectrum.  The pseudo decay-time $t_z$ is defined as

\begin{equation}
t_z  \equiv \frac{ (z_{\rm decay} - z_{\rm PV}) \times M } { p_{z} },
\end{equation}

\noindent where $z_{\rm decay} - z_{\rm PV}$ is the difference between the positions of the reconstructed vertex of the $\jpsi \pip \pim$ and the collision vertex along the beam axis, $M$ is the known mass\cite{PDG2018} of the reconstructed $\psitwos$ or $\X$ candidate, and $p_z$ is the candidate's momentum along the beam axis.  The signal in the $t_z$ spectrum is fit with a delta function representing the prompt component and an exponential decay function representing the component from $b$ decays, which are convolved with a double Gaussian resolution function.  Two different parameterizations of the $t_z$ background components using mass sidebands above and below the mass peak of interest are employed. The first is an empirically determined analytical function as was done in Ref.~\cite{LHCb-PAPER-2018-049} and the second directly uses the $t_{z}$ shape templates taken from the mass sidebands in the data.  

\begin{figure}[tbp]
  \centering
  \includegraphics[width=1.0\linewidth]{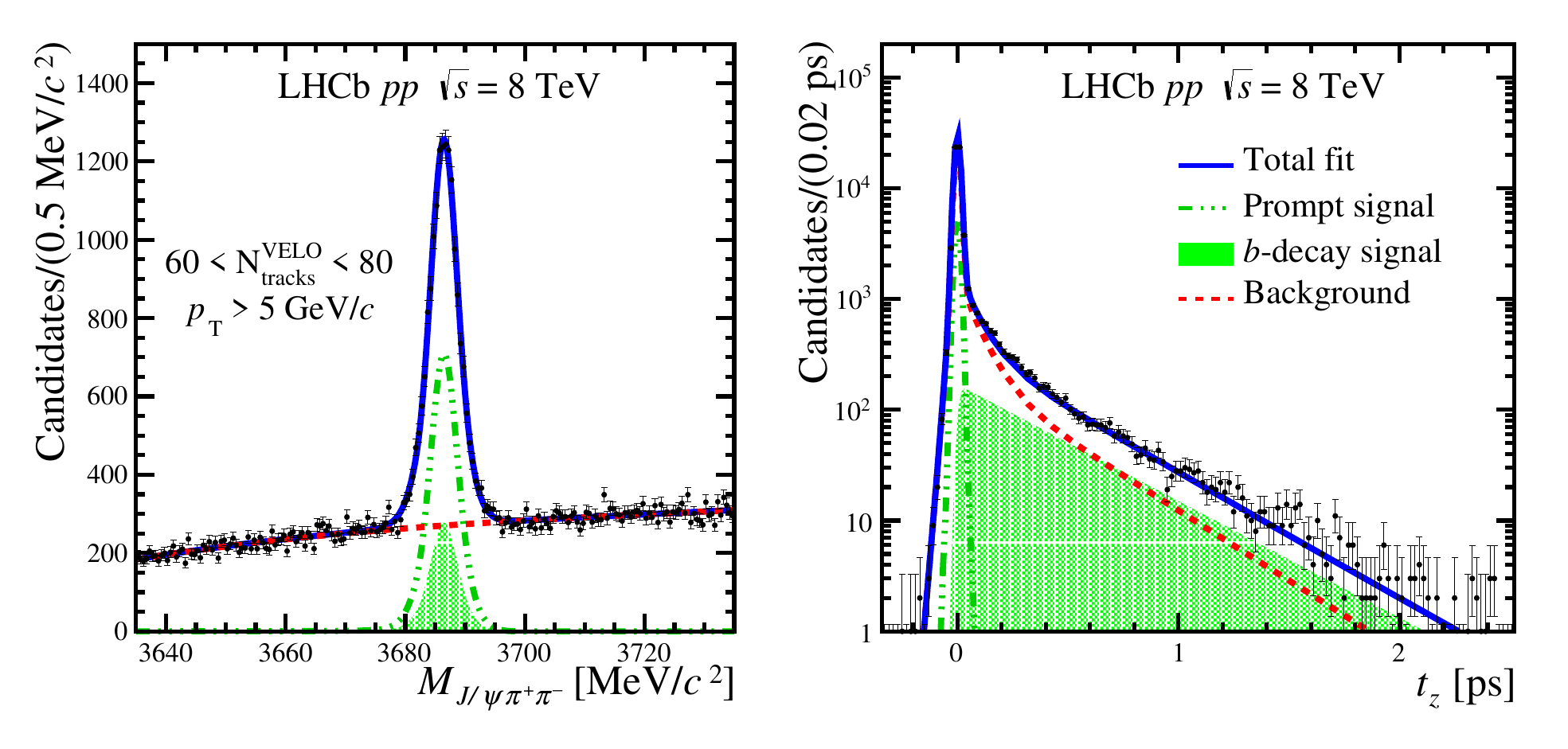}
  \caption{The $\psitwos$ (left) invariant-mass and (right) $t_z$ spectrum in the $\pt$ and multiplicity ranges $\pt> 5 \gevc$ and $60 <\N < 80$, with the simultaneous fit superimposed.}
  \label{fig:newFig}
\end{figure}

In the fit to the invariant-mass spectrum, the $\psitwos$ peak is represented by a sum of two Crystal Ball functions, as in a previous LHCb analysis at $7 \tev$ \cite{LHCb-PAPER-2011-045}.  The measured $\X$ peak is well described by a Gaussian function.  The background contribution is studied by examining the invariant-mass spectrum constructed by like-sign pion pairs, and is found to be well described by third-order Chebyshev polynomials; this shape is used to represent the background when fitting the $\jpsi\pip\pim$ mass spectra.  The invariant mass and $t_{z}$ spectra are divided into bins of $\N$, and the fit is performed in each bin, separately for the $\psitwos$ meson and $\X$ state.  An example is shown in Fig.~\ref{fig:newFig}.

\begin{figure}[tbp]
  \centering
  \includegraphics[width=1.0\linewidth]{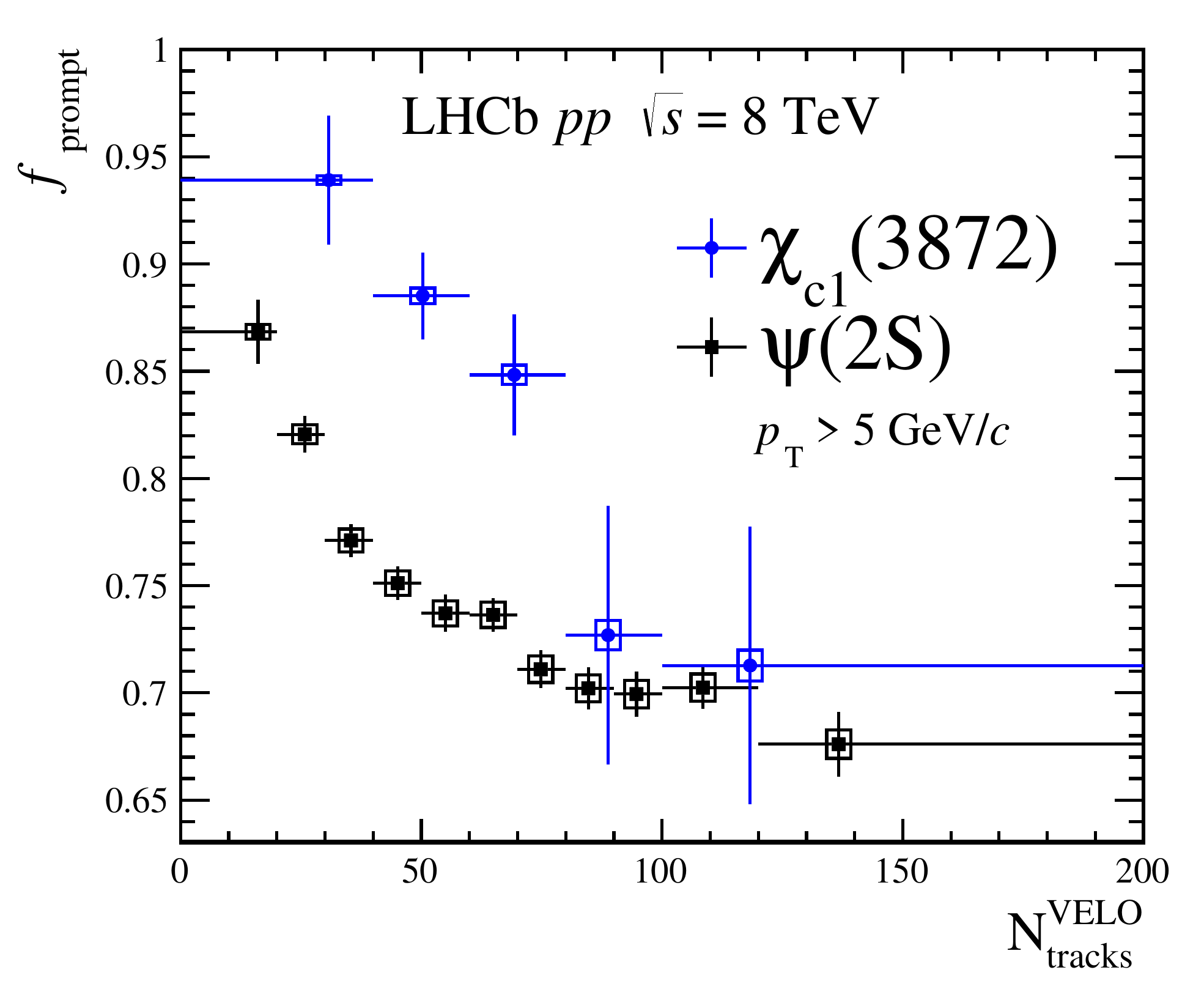}
  \caption{The fraction, \fprompt, of promptly produced $\X$ and $\psitwos$ hadrons, as a function of the number of tracks reconstructed in the VELO.  The vertical error bars (boxes) represent the uncorrelated (correlated) uncertainties, while the horizontal error bars indicate bin widths.}
  \label{fig:f_prompt}
\end{figure}

The total yield and the measured fraction of the inclusive signal that is produced at the collision vertex for each state is determined by a fit. Due to the different production mechanisms, the observed $\X$ and $\psitwos$ hadrons from these sources have different momentum distributions, which may lead to differences in acceptance and reconstruction efficiencies.  To account for this effect, the $\pt$ distributions of prompt and displaced signal candidates are extracted from the data using the $sPlot$ technique \cite{splot}, and the $\pt$ distributions of the simulated particles are weighted to match those of the data.  Since these measurements are binned in multiplicity and effects of multiplicity-dependent breakup may depend on $\pt$, the simulation is reweighted to match $\pt$ distributions extracted from low- and high-multiplicity samples.  The difference in the acceptance and reconstruction efficiencies found using these different parameterizations of the $\pt$ distributions is taken as a systematic uncertainty.   Corrections are applied to account for the relative acceptance of the LHCb spectrometer between particles produced at the primary vertex and in $\bquark$ decays  $\ensuremath{\varepsilon_{\mathrm{prompt}}^{\mathrm{acc}}/\varepsilon_{\mathrm{\bquark}}^{\mathrm{acc}}}\xspace$, which is determined via simulation to be 1.00$\pm$0.01 for the $\psitwos$ state and 1.02$\pm$0.01 for the $\X$ state, and for the relative reconstruction and selection efficiencies $\ensuremath{\varepsilon_{\mathrm{prompt}}^{\mathrm{reco}}/\varepsilon_{\mathrm{\bquark}}^{\mathrm{reco}}}\xspace$, which are 0.99$\pm$0.03 for the $\psitwos$ state and 1.11$\pm$0.04 for the $\X$ state.  The central value of $\fprompt$ is taken as the average of the values obtained from the two fitting methods, while the difference is taken as a systematic uncertainty, which ranges from 1$\%$ to 2$\%$ for the $\psitwos$ fits and from 2$\%$ to 6$\%$ for the $\X$ fits.  This uncertainty is uncorrelated between bins and is comparable to the statistical uncertainty on the $\psi(2S)$ data, while the statistical uncertainty dominates on the $\X$ data.  The uncertainty on the relative efficiency is taken as a systematic uncertainty on $\fprompt$, which is correlated between the data points for each species.  The resulting values of \fprompt as a function of multiplicity are shown in Fig.~\ref{fig:f_prompt}, up to \mbox{$\N$ = 200}. The fraction of events with $\N>$200 is negligible and is not included in the analysis. The horizontal position of each point is the average value of $\N$ for signal events within that bin. 

A clear decrease of $\fprompt$ is seen as the multiplicity increases, for both the $\psitwos$ and $\X$ hadrons. This could be due to a combination of several effects: the average multiplicity is higher in events containing a $\bquark\bar{\bquark}$ pair due to their fragmentation into hadrons and subsequent decays \cite{Rowson:1985xh,Abbiendi:2002vn}; or the suppression of prompt $\psitwos$ and $\X$ production via interactions with other particles produced at the vertex, which decreases the prompt production in high-multiplicity events, but does not affect production in $b$ decays. 

 The prompt and $b$-decay components are examined directly by calculating the ratio of the $\X$ and $\psitwos$ cross-sections, $\sigma_{\chi}/\sigma_{\psi}$, times their respective branching fractions to the $ \jpsi \pip \pim$ final state, $\BR_\chi$ and $\BR_\psi$.  This ratio is given by



\begin{equation}
\label{eqn:rat}
\begin{aligned}
\frac{\sigma_{\chi}}{\sigma_{\psi}}  \frac{\BR_\chi}{\BR_\psi}=\frac{N_{\chi} \, \fprompt^{\chi}}{N_{\psi} \, \fprompt^{\psi}} \frac{\varepsilon_{\psi}^{\mathrm{acc}}}{\varepsilon_{\chi}^{\mathrm{acc}}} \frac{\varepsilon_{\psi}^{\mathrm{reco}}}{\varepsilon_{\chi}^{\mathrm{reco}}}  \frac{\varepsilon_{\psi}^{\mathrm{PID}}}{\varepsilon_{\chi}^{\mathrm{PID}}} .\\
\end{aligned}
\end{equation}
\noindent Here, $N$ is the signal yield, $\fprompt$ is the prompt fraction and the $\varepsilon$ terms represent various efficiency corrections of the corresponding state. The ratio of cross-sections from $\bquark$ decays is found by replacing $\fprompt$ with (1$-\fprompt$) in Eq.~\ref{eqn:rat}. Correlated systematic uncertainties largely cancel in the ratio, and the result is dominated by uncorrelated uncertainties. The ratio of efficiencies for four charged decay products to fall within the LHCb acceptance, $\ensuremath{\varepsilon_{\psi}^{\mathrm{acc}}/\varepsilon_{\chi}^{\mathrm{acc}}}\xspace$, is found via simulation to be consistent with one with an uncertainty of approximately 1$\%$ that is determined by varying the $\pt$ distributions of the simulated $\psitwos$ and $\X$ hadrons. Control samples of identified muons and pions obtained from data are used to measure the ratio of muon and pion identification efficiencies,  $\ensuremath{\varepsilon_{\psi}^{\mathrm{PID}}/\varepsilon_{\chi}^{\mathrm{PID}}}\xspace$, which is near one with an uncertainty of about 1$\%$ due to the finite size of the control sample.  The only relative efficiency that has a significant deviation from unity is the ratio of reconstruction efficiencies, $\ensuremath{\varepsilon_{\psi}^{\mathrm{reco}}/\varepsilon_{\chi}^{\mathrm{reco}}}\xspace$, which is found via simulation to be 0.58$\pm$0.02 (0.65$\pm$0.04) for particles that are produced promptly (in $\bquark$-decays).  This is due to the different kinematic properties of the pion pair produced in the decays: pions from $\X$ hadron decays proceed through an intermediate $\rho^{0}(770)$ resonance \cite{Abulencia:2005zc} and have a higher reconstruction efficiency than pions from the $\psitwos$ decay due to their higher $p_{T}$.  The uncertainty on the ratio of reconstruction efficiencies is taken from the variations observed when weighting the $\pt$ distributions of the simulated $\psitwos$ and $\X$ hadrons to match those in the data in different multiplicity bins, as previously discussed.

\begin{figure}[tbp]
  \centering
  \includegraphics[width=1.0\linewidth]{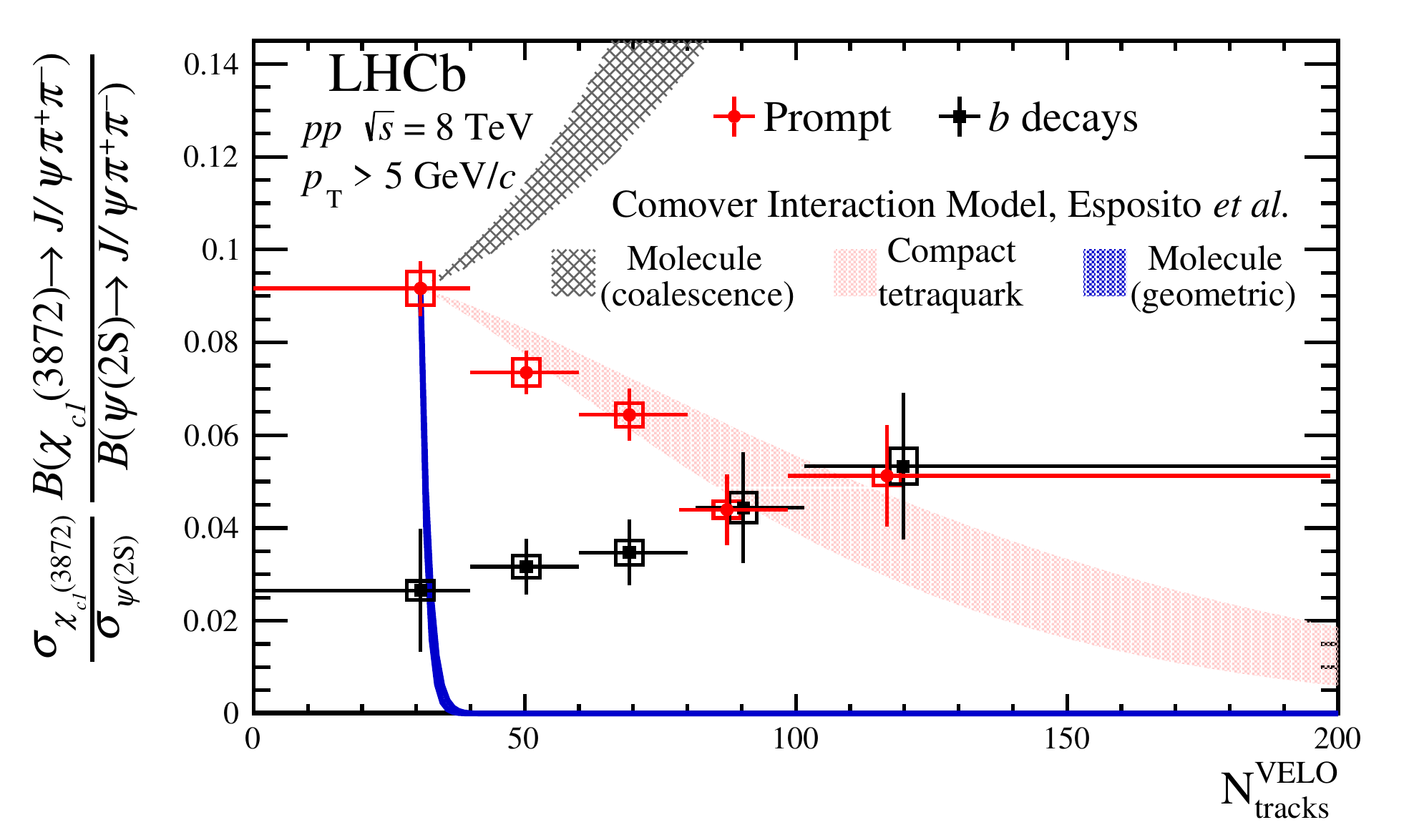}
  \caption{The ratio of the $\X$ and $\psitwos$ cross-sections measured in the $\jpsi \pip \pim$ channel as a function of the number of tracks reconstructed in the VELO.  The point-to-point uncorrelated (correlated) uncertainties are shown as vertical error bars (boxes) and the bin widths are shown as horizontal error bars.  See text for details on calculations from Ref.~\cite{Esposito:2020ywk}}
  \label{fig:rat}
\end{figure}

The ratio of cross-sections is shown in Fig.~\ref{fig:rat}. A decrease in the prompt production of $\X$ hadrons relative to prompt $\psi(2S)$ mesons is observed as the charged particle multiplicity increases.  To illustrate this effect, a linear fit to this data, which considers only the uncorrelated uncertainties, is performed and returns a negative slope that differs from zero by 5 standard deviations.  

After preliminary LHCb results on multiplicity-dependent $\X$ production were presented \cite{QMproc}, calculations of these observables based on the Comover Interaction Model \cite{Capella:2000zp, Ferreiro_comovers} were performed \cite{Esposito:2020ywk}.  In this model, promptly produced $\X$ and $\psitwos$ hadrons interact with other produced particles, with a breakup cross-section $\sigma_{\rm br}$ that is determined by their radius and binding energy.  The model assumes no interactions at low multiplicity, and the calculations are normalized to the data in the lowest multiplicity bin.  A purely molecular $\X$ has a large radius and correspondingly high $\sigma_{\rm br}$, and is quickly dissociated as multiplicity increases.  If coalescence provides an additional formation mechanism for molecular $\X$, the ratio $\sigma_{\X}/\sigma_{\psitwos}$ rises with multiplicity.  Neither of these calculations are consistent with the data.  A compact tetraquark $\X$ has a slightly larger radius and $\sigma_{\rm br}$ than the $\psitwos$, and in this scenario $\sigma_{\X}/\sigma_{\psitwos}$ gradually decreases with multiplicity, matching the measured trend.

In contrast to the prompt data, the ratio of cross-sections for production in $\bquark$ decays shows a slight increase, which is not statistically significant.  A linear fit to these data points, again without considering the correlated systematic uncertainty, gives a positive slope that is consistent with zero within 1.6 standard deviations.  Since these hadrons originate from displaced decay vertices of $\bquark$ hadrons, they are not subject to suppression via interactions with other particles produced at the primary vertex.  Consequently, this ratio is set only by the branching fractions of $\bquark$ decays to $\X$ and $\psitwos$ hadrons.  The multiplicity dependence of $b$ hadron production has not been studied in detail and modification of the $\bquark$ hadron admixture could affect $\X$ production, as different $\bquark$ hadron species may have different decay probabilities to $\X$ states \cite{Aaboud:2016vzw,Sirunyan:2020qir}. However, the uncertainties preclude drawing any firm conclusions on multiplicity-dependent modifications of $\bquark$ hadronization from this data.

In conclusion, the prompt $\X$ and prompt $\psitwos$ production cross-sections decrease relative to their production via $b$ decays as the charged particle multiplicity increases in $pp$ collisions at $8\tev$.  A comparison between the $\X$ and $\psitwos$ states shows that, in contrast to production from $b$-decays, which display no significant dependence on multiplicity, prompt production of $\X$ is suppressed relative to prompt $\psitwos$ production as multiplicity increases.  This observation is an important ingredient to obtain
a full understanding of the nature of the $\X$ state.

\section*{Acknowledgements}
%
%
\noindent We express our gratitude to our colleagues in the CERN
accelerator departments for the excellent performance of the LHC. We
thank the technical and administrative staff at the LHCb
institutes.
We acknowledge support from CERN and from the national agencies:
CAPES, CNPq, FAPERJ and FINEP (Brazil); 
MOST and NSFC (China); 
CNRS/IN2P3 (France); 
BMBF, DFG and MPG (Germany); 
INFN (Italy); 
NWO (Netherlands); 
MNiSW and NCN (Poland); 
MEN/IFA (Romania); 
MSHE (Russia); 
MICINN (Spain); 
SNSF and SER (Switzerland); 
NASU (Ukraine); 
STFC (United Kingdom); 
DOE NP and NSF (USA).
We acknowledge the computing resources that are provided by CERN, IN2P3
(France), KIT and DESY (Germany), INFN (Italy), SURF (Netherlands),
PIC (Spain), GridPP (United Kingdom), RRCKI and Yandex
LLC (Russia), CSCS (Switzerland), IFIN-HH (Romania), CBPF (Brazil),
PL-GRID (Poland) and OSC (USA).
We are indebted to the communities behind the multiple open-source
software packages on which we depend.
Individual groups or members have received support from
AvH Foundation (Germany);
EPLANET, Marie Sk\l{}odowska-Curie Actions and ERC (European Union);
A*MIDEX, ANR, Labex P2IO and OCEVU, and R\'{e}gion Auvergne-Rh\^{o}ne-Alpes (France);
Key Research Program of Frontier Sciences of CAS, CAS PIFI,
Thousand Talents Program, and Sci. \& Tech. Program of Guangzhou (China);
RFBR, RSF and Yandex LLC (Russia);
GVA, XuntaGal and GENCAT (Spain);
the Royal Society
and the Leverhulme Trust (United Kingdom).




\addcontentsline{toc}{section}{References}
\bibliographystyle{LHCb}
\bibliography{main,standard,LHCb-PAPER,LHCb-CONF,LHCb-DP,LHCb-TDR}

\newpage
\centerline
{\large\bf LHCb collaboration}
\begin
{flushleft}
\small
R.~Aaij$^{31}$,
C.~Abell{\'a}n~Beteta$^{49}$,
T.~Ackernley$^{59}$,
B.~Adeva$^{45}$,
M.~Adinolfi$^{53}$,
H.~Afsharnia$^{9}$,
C.A.~Aidala$^{84}$,
S.~Aiola$^{25}$,
Z.~Ajaltouni$^{9}$,
S.~Akar$^{64}$
% ...

{\footnotesize \it

% ...
$ ^{9}$Universit{\'e} Clermont Auvergne, CNRS/IN2P3, LPC, Clermont-Ferrand, France\\
% ...
$ ^{25}$INFN Sezione di Milano, Milano, Italy\\
% ...
$ ^{31}$Nikhef National Institute for Subatomic Physics, Amsterdam, Netherlands\\
% ...
$ ^{45}$Instituto Galego de F{\'\i}sica de Altas Enerx{\'\i}as (IGFAE), Universidade de Santiago de Compostela, Santiago de Compostela, Spain\\
% ...
$ ^{49}$Physik-Institut, Universit{\"a}t Z{\"u}rich, Z{\"u}rich, Switzerland\\
% ...
$ ^{53}$H.H. Wills Physics Laboratory, University of Bristol, Bristol, United Kingdom\\
% ...
$ ^{59}$Oliver Lodge Laboratory, University of Liverpool, Liverpool, United Kingdom\\
% ...
$ ^{64}$University of Cincinnati, Cincinnati, OH, United States\\
% ...
$ ^{67}$Syracuse University, Syracuse, NY, United States\\
% ...
$ ^{84}$University of Michigan, Ann Arbor, United States, associated to $^{67}$\\
% ...
\medskip
}
\end{flushleft}

\end{document}